\documentclass{aastex63}
\usepackage{epsfig}
\usepackage{amsmath}
\usepackage{verbatim}

\shorttitle{long lived technological civilizations}
\shortauthors{B. Zuckerman}

\begin{document}

\title{Infrared and Optical Detectability of Dyson Spheres at White Dwarf Stars}

\correspondingauthor{B. Zuckerman} 
\email{ben@astro.ucla.edu}

\author[0000-0001-6809-3045]{B. Zuckerman}
\affiliation{Department of Physics and Astronomy, University of California, Los An
geles, CA 90095-1562, USA}


\begin{abstract}

It has been hypothesized that advanced technological civilizations will construct giant space colonies
and supporting infrastructures to orbit about their home stars.   With data from recent satellites that operate at infrared and optical wavelengths (Spitzer, WISE, TESS, Kepler), in company with a few modest assumptions, it is now possible to begin to constrain observationally the frequency of such space-based civilizations in our Milky Way Galaxy.


\end{abstract}

\section{Introduction}

The number of technological civilizations in our Milky Way Galaxy is a topic of
endless debate and speculation.  It is generally agreed that, if such civilizations
are abundant, then they must be long lived.   Thus, such civilizations will ultimately have to deal with the
evolution of their stars, first into red giants and then into white dwarfs.  Two
possible paths would be mass migration to another star (Zuckerman 1985; Hansen \& 
Zuckerman 2021) or
remaining at the evolving star and accommodating to the changes in luminosity and
stellar wind flux (Gertz 2020).  These are not mutually exclusive and both paths
could be taken by an advanced civilization.


One plausible picture of an advanced civilization that accommodates stellar
evolution -- rather than entirely fleeing from it -- would include numerous
giant space colonies and other giant constructs in orbit in sphere- or
ring-like configurations about a white dwarf or an old main sequence star.  This type of technological
civilization is often referred to as a Dyson sphere or ring, after Freeman Dyson who
first described it (Dyson 1960). One motivation for giant constructs would be
to supply energy to the space colonies, especially those in orbit around white dwarfs, because white dwarfs are of low and declining luminosity.  

Whatever the detailed arrangement, one anticipates that energy
collection devices would be situated close to stars so as to minimize
their size.  Their temperatures would probably lie in the range 300
to 1000 K and they would radiate primarily between a few and 10 $\mu$m wavelengths
(Wright et al 2014).  This emission would be in addition to (an excess above) emission
from the star itself.  It seems unlikely that carbon/water based life
could ever exist at temperatures much above 300 K. But should such life evolve
(redesign itself) into something very different, then perhaps the space colonies
themselves would reside at temperatures substantially above 300 K.

The focus of the present paper is the detectability of artificial constructs in orbit around white dwarf stars, although main sequence stars are also considered (Section 6).   The past two decades have witnessed the birth of the study of planetary systems around white dwarfs via detection of elements heavier than helium in their photospheres, detection of infrared emission from orbiting dust particles and optical spectral lines from orbiting gas, and occasional discovery of orbiting planets and brown dwarfs via either their IR emission or a transit in front of a white dwarf.   The photospheric heavy elements are seen in at least 25\% of all white dwarfs and is a marker for the presence of orbiting dust particles and gas.  The dust can reveal itself directly via excess IR emission above that emitted by the white dwarf photosphere, but such excess is seen in only a few percent of all white dwarfs (Rocchetto et
al. 2015, Farihi 2016, Wilson et al. 2019).  Even less common is excess IR emission emitted by a cool brown dwarf or cool white dwarf.  Thus, in order to identify a Dyson sphere/ring at a white dwarf, one must eliminate dust particles and brown and white dwarfs as the carrier of excess IR emission.

In addition to absorbing radiation emitted b‌y the white dwarf, the space constructs will 
sometimes pass (transit) between the star and our telescopes, thus causing a dip in the 
received optical light. Arnold (2005) discussed the transit light-curve signatures of very large artificial objects.   Agol (2011) considered transits of Earths in the habitable zones
of white dwarfs.  The first examples of cool obects transiting white dwarfs have recently
been obtained with the Zwicky Transient Facility and NASA's TESS satellite (Vanderbosch et al 2020 \& 2021; Vanderburg et al 2020; van Roestel et al 2021). 

Zuckerman (1985) showed that a few percent of the technological civilizations in
our Milky Way galaxy may already have experienced the evolution of their star
from main sequence to white dwarf (see Section 4
below).  Thus, if technological civilizations are common, as some believe,
then finding one in orbit around a white dwarf could be quite plausible.

In the following we refer to Dyson spheres/rings as DSRs. The primary purpose of
the present paper is to consider and compare the observational limits that can currently be
placed on the number (frequency) of DSRs at white dwarf stars via their infrared emission and/or
transit frequency.  Many guesstimates for the number of technological 
civilizations in the Milky Way can be found in the literature (see Section 2).  But now, thanks
to advances in discovery and analysis of extrasolar planetary systems, it is possible to
frame the question in a more quantitative way:  On what fraction of planets in the habitable zone
of sun-like stars do long-lived technological civilizations arise and eventually build a DSR?

In the following section we define what is meant by the number of technological
civilizations in our Galaxy.  In Section 3 we consider the carriers of excess IR
emission at white dwarf stars.  Section 4 is a discussion of how infrared bright
a DSR would have to be to have already been detected at a white dwarf star and what constraints such measurements place on the frequency of DSRs. Section 5 is a consideration of transit detectability of DSRs and how it compares with detection via infrared emission.  Section 6 describes
the current observational situation at main sequence
stars.  Section 7 compares the advantages and disadvantages of white dwarfs compared to main sequence stars when one is searching for a DSR.

\section{What is meant by the number of technological civilizations in the Galaxy?}

In order to give context to limits on the frequency of DSRs in the Galaxy, one should have some idea of how many long-lived technological civilizations “N” might be anticipated.   The range suggested in the literature is huge, extending from zero (Hart 1975) to as many as 10 billion (F. Drake as quoted in Tarter 2001).   Other estimates include: 50,000 to 1,000,000 (Shklovskii \& Sagan 1966), one billion (Oliver \& Billingham 1971, Goldsmith \& Owen 1992), and 10 million (Sagan 1980).   As a consequence of the discovery during the past few decades that there are more planets than stars in the Milky Way, many persons would likely expect a (very) large value for N.

In each of the above examples, N refers to the number of independently arising technologies and does not include the possibility that a technological civilization might expand to occupy one or more additional star systems.   The relevance for the present paper of any such migrations depends strongly on the motivations for migration.   

If the only motivation for interstellar travel is to escape from one’s home planetary system so as to avoid the evolution of the home star to red giant and white dwarf (Zuckerman 1985, Hansen \& Zuckerman 2021), then one would never expect a technological civilization to build a DSR around any white dwarf other than the white dwarf that their home star evolves into.  Because it is hard to imagine any reason why a technological civilization would want to migrate to a white dwarf and then construct a DSR, it is probably safe to assume that the frequency of DSRs at white dwarf stars will depend only on the number of independently arising technological civilizations. In any event, in the unlikely event that migration to white dwarfs does occur frequently, then this would only serve to increase the likelihood of finding a DSR at a white dwarf.

By contrast, interstellar travel and colonization might increase the number of DSRs that exist at main sequence stars.   As one example, if curiosity about biology on other worlds is an important motivation for interstellar travel (e.g., Zuckerman 2019), then some DSRs at main sequence stars might be constructed by a technological civilization that first arose at a different star.  

In the discussion to follow we assume that N is equal to the number of independently arising technological civilizations in the Milky Way.   We want to estimate what constraints existing and potential future infrared and optical observations of white dwarf and main sequence stars place on the frequency of DSRs and thus on N.

\section{White dwarfs with excess infrared emission}

A few percent of white dwarf stars display excess infrared emission – characteristic temperatures $\sim$1000 K -- above their photospheres (e.g., Rocchetto et al 2015).  From the first discovery, such emission has generally been attributed to either orbiting dust particles or an orbiting brown dwarf (Zuckerman \& Becklin 1987).    It is now understood that the carrier of the IR emission is almost always orbiting dust particles, with but few examples of (spatially unresolved) brown dwarf or cool white dwarf companions (e.g., see Introduction in van Roestal et al 2021).  A summary of studies of IR emission from white dwarfs can be found in the Introduction and in Table 1 in Xu et al (2020).  
When dust is responsible for an IR excess, then, with no reliably known counter examples, “pollution” of the white dwarf photosphere by elements heavier than helium is also observed.    The generally accepted model is that the material contained in the dust particles is accreted onto the white dwarf (Veras 2016 \& 2021; Farihi 2016; Jura \& Young 2014, Zuckerman \& Young 2018).  The resulting level of photospheric pollution for stars with excess IR emission is typically large with respect to the average pollution level seen in all polluted white dwarfs.   The parent body for the dust is usually a rocky asteroid or asteroids (Jura 2003 \& 2008) that have been torn apart by the strong gravitational field of the white dwarf.

Thus, if heavy elements are detected in the photosphere of a white dwarf with excess IR emission, then it is likely that the excess is due to dust particles and not to a brown dwarf or a DSR.   Because excess IR emission carried by dust particles is less likely for cool (cooling times $>$1000 Myr) white dwarfs (e.g., Table 3 in Rocchetto et al 2015), statistically, white dwarfs with excess IR emission and relatively low temperatures $<$8000 K could more likely host a DSR.  But some white dwarfs with temperatures as low as $\sim$6000 K (Debes et al 2019; Gentile Fusillo et al 2019) and 4200 K (Hollands et al 2021) appear to host excess IR emission likely due to dust.
A typical blackbody temperature of the dust at white dwarfs is $\sim$1000 K, but a few are known with temperatures as low as 300-400 K (Table 3 in Rocchetto et al 2015).  As mentioned above, these are plausible temperatures for a DSR.  

WD2328+107 probably has an IR excess (Wilson et al 2019) and has a progenitor main sequence mass similar to that of the Sun (Rocchetto et al 2015), but no evidence for heavy elements in its atmosphere.  Still, the putative excess IR emission might be carried by a brown dwarf or a cool white dwarf companion (Wilson et al 2019).  A JWST observation should be able to clarify the situation. 


Based on the above considerations, with the possible exception of WD2328+017, it seems unlikely that any currently known example of excess IR emission would be a good DSR candidate. Elimination of dust as the carrier of IR emission requires, at a minimum, the absence of photospheric pollution and silicate emission features in the 10 $\mu$m spectrum; such features are usually present (Jura et al 2009) when dust is present.   

\section{The Infrared Detectability of Dyson Spheres at white dwarf stars}

What limits can currently be placed on the frequency of DSRs at white dwarfs?  While it might be possible to discover a narrowband signal generated by an extraterrestrial intelligence (ETI) from a white dwarf system simply by accident, there have not been any published radio or optical searches of white dwarfs meant specifically to detect ETI.  Thus, the limits on N described below are the first to be obtained for white dwarfs.   

We frame the following discussion in two ways.  In the first, we estimate upper limits to N that are implied by Spitzer Space Telescope surveys of white dwarfs in two limits: (1) all N technological civilizations in the Galaxy arise on stars with masses similar to that of the Sun (see Equation 2), and (2) technological civilizations arise with equal probability on stars with masses less than or about equal to that of the Sun.  A second way of approaching the matter does not involve N directly, but rather asks the question: if a certain fraction, $\alpha$, of G-type stars in the Galaxy contain an appropriate rocky planet in the habitable zone such that life might originate, then what fraction of these potentially habitable planets eventually evolve a technological civilization that builds a DSR?  Estimates for $\alpha$ have been derived from Kepler data (see below).

Zuckerman (1985) calculated the number of technological civilizations that have originated on planets that orbit around main sequence stars that have evolved to white dwarfs in a time that is less than the age of our Galaxy which he, and we, assume to be 10$^{10}$ years.   We redo these calculations here, but using more recent information. 

Sollima (2019) considers a variety of star formation histories.   For the purposes of the present paper, assumption of a constant birthrate during the past 10$^{10}$
years is adequate.    For stars with masses about equal to and greater than a solar mass, following Sollima, we assume that dQ/dt, the rate of formation of stars in the Galaxy with masses between M and M+dM, is given by    :

\begin{equation}
\frac{dQ}{dt} \ = \ 1.55 \frac{dM}{M^{2.55}}
\end{equation}

In the above and below expressions, Q is simply a number, but we cannot label it with ``N'' because historically and, as above, N has been reserved to denote the number of technological civilizations in the Galaxy The exponent 2.55 is the mean of the range of power law slopes, 2.4-2.7, suggested by Sollima.
And, following Iben (1984), we have normalized to a Galactic birthrate of one star per year with a mass greater than M$_{\odot}$.   Sollima gives a history of the evolution of the power law exponent going back
to Salpeter (1955) who adopted 2.35 over the range of stellar masses of interest here.

The white dwarf initial-final mass relation for progenitor stars with main sequence masses between 0.85 and 7.5 M$_{\odot}$ is given in Cummings et al (2018).   Mamajek (2021)\footnote{http://www.pas.rochester.edu/$\sim$emamajek/EEM\_dwarf\_UBVIJHK\_colors\_Teff.txt}
has compiled an extensive list of the properties of main sequence stars, in particular, spectral type vs. stellar luminosity and mass.  We assume that stars with main sequence lifetimes that  are shorter than 4.5 x $10^9$ years  (M $>$1.25 M$_{\odot}$) do not last sufficiently long to permit a 
technological civilization to evolve on an orbiting planet.   And we assume a minimum main sequence mass of 0.95 M$_{\odot}$ to allow some time for early production of metals with which to make rocky 
planets.  Therefore, the interesting range of stellar masses lies between 0.95 and 1.25 M$_{\odot}$.  According to Mamajek, these masses correspond to spectral types G7 and F6, respectively, and with luminosities 0.74 and 2.69 L$_{\odot}$  Thus, stars with nearly the mass of the Sun have main-sequence lifetimes that vary like M$^{-3.8}$.

The number of stars in the Milky Way born in the mass range 0.95 to 1.25 M$_{\odot}$ during the past $10^{10}$ years is:

\begin{equation}
Q_{\rm i} \ = \ 10^{10}\int_{0.95}^{1.25} \frac{1.55 \, dM}{M^{2.55}} \ = \ 3.75 \times 10^{9} \, {\rm stars.}
\end{equation}

The number of these stars still on the main sequence is:

\begin{equation}
Q_{\rm a} \ = \ \int_{0.95}^{1.25}1.55 \,  \frac{dM}{M^{2.55}} \int_{0}^{10^{10}/M^
{3.8}}dt \ = \ 2.95 \times 10^{9} \, {\rm stars.}
\end{equation}

So the number of stars that have “died” and become white dwarfs is:

\begin{equation}
Q_{\rm i} - Q_{\rm a} \ = \ 8 \times 10^{8}
\end{equation}

Now we want to estimate how many technological civilizations (N*) might have existed on planets orbiting these 8 x $10^8$ white dwarfs.  Let’s first assume that all N technological civilizations in the Galaxy originated around stars in the main sequence mass range 0.95 to 1.25 M$_{\odot}$.  Then

\vskip 0.2in

                         N*$_{max}$ = (8 x $10^8$)N/(3.75 x $10^9$) = 0.2 N

\vskip 0.2in


If, for example, N = $10^9$,  so N*$_{max}$ = 2 x $10^8$ is distributed among 8 x $10^8$ white dwarfs.  Then 25\% of all white dwarfs could potentially display a DSR.   

How many white dwarfs have been searched for excess IR emission and to what level of sensitivity?  And how many of these had masses on the main sequence in the range 0.95 to 1.25 M$_{\odot}$?
The relevant satellites are Spitzer and the Wide-field Infrared Survey Explorer (WISE).
Various published papers describe Spitzer surveys for excess IR emission of white dwarfs.    
Table A1 in Rocchetto et al (2015) lists 134 white dwarfs of which 30 had progenitors with main sequence masses $\leq$1.25 M$_{\odot}$.  Wilson et al. (2019) carried out a more extensive survey (of 217 white dwarfs) that included the Rocchetto stars as a subset.  Assuming the same fraction of 
low mass progentors among the 217 stars as among the 134, yields a total of about 50 white dwarfs in the relevant mass range.   Figure 10 in Farihi (2016) shows 108 white dwarfs with photospheres that are polluted with heavy elements that were searched with Spitzer for IR excess emission during its first 7 cycles.  Most of these stars are cooler than the stars in the Rocchetto/Wilson sample, so there is little overlap.  Wilson et al note that only about one in 30 white dwarfs with polluted photospheres exhibits an IR excess in 3-4 $\mu$m IRAC photometry.  Putting all of these considerations together, suggests that the Farihi sample of 108 stars contains about 25 that lack  detectable excess IR emission and had main sequence masses that fall in the relevant range (less than 1.25 M$_{\odot}$).   Yet another major white dwarf survey was by Mullally et al (2007), with $\sim$100 searched.   Finally there were some smaller Spitzer surveys, for example by Farihi et al (2008) of 17 white dwarfs.

In total, it appears that Spitzer surveyed at least 100 white dwarfs that had masses on the main sequence in the range 0.95 to 1.25 M$_{\odot}$ and with no evidence for excess IR emission.   Table 3 in Rocchetto et al (2015) indicates that the typical search sensitivity -- excess IR luminosity divided by bolometric luminosity -- was about 0.1\%, for excess emission at temperatures between about 300 and 1700 K.   Thus, the considerations that follow apply only for DSRs of fractional IR luminosity greater than 0.1\%.

As noted just above, if all technological civilizations originated on stars with main sequence spectral types between G7 and F6, and if N is equal to one billion, then about 25\% of the 100 Spitzer-observed white dwarfs with these spectral types and without orbiting dust or a cool companion could potentially display a DSR.  Yet none have been reported in the literature where an IR excess exists that could not definitely or probably be attributed to dust or to a companion. 

If N = 100 million, then we might have expected to detect 2.5 DSRs among the 100 relevant Spitzer white dwarfs.  Given statistics of small numbers, these few DSRs might not be included among the 100 observed white dwarfs.  So, for the situation where all technological civilizations in the Milky Way orginate on main sequence stars with spectral types betwen G7 and F6, and where a DSR around the white dwarf that these stars evolve into has a fractional excess IR luminosity of at least 0.1\%, then the implication is that N is less than 100 million.

Based on the IMF for stellar masses less than a solar mass deduced by Sollima (2019), G7 through F6 stars comprise about 10\% of all stars with main sequence lifetimes that are long enough to allow evolution of life to technology.  If we assume that the N technological civilizations arise with equal probability around all such stars with no preference for stellar mass, then the implication from white dwarfs observed with Spitzer is that N is less than $10^9$.

In summary, Spitzer observations of white dwarfs indicate that the upper limit to N is about 300 million, plus or minus  a few hundred million; the better ($<$$10^8$) and poorer ($<$$10^9$) bounds to N apply, respectively, in the cases where technological civilizations preferentially form on planets at G-type stars or, alternatively  if they form with equal probabilites on planets that orbit stars with masses less than or equal to that of the Sun.  Again, the technological civilization must have contructed a DSR with fractional IR luminosity of at least 0.1\% for these limits to apply (see Section 7 for consideration of whether this is a plausible DSR luminosity). 

From Equation 4 we see that about a billion F6 through G7 stars that were on the main sequence are now white dwarfs.  Studies of Kepler and other data bases by Zink \& Hansen (2019) and by Bryson et al. (2021) suggest that about 30\% of G-type stars are orbited by a potentially habitable planet, or about 300 million such planets that orbit the white dwarfs of interest here.  If as many as one in 30 of these planets spawns life that eventually evolves to a state where it constructs a DSR with luminosity at least 0.1\% that of its host white dwarf, then in a sample of 100 white dwarfs we might have expected to see a DSR.  Thus, fewer than 3\% of the habitable planets that orbit sun-like stars host life that evolves to technology, survives to the white dwarf stage of stellar evolution, and builds a DSR with fractional IR luminosity of at least 0.1\%.

The WISE mid-infrared survey of the sky (Wright et al 2010) was sufficiently sensitive to yield significant limits for N for main sequence stars (see Section 6).  But for the much fainter white dwarfs, clear identification of excess IR emission is plagued by confusion (e.g., Dennihy et al 2020; Xu et al 2020) with other sources of IR emission.  Hopefully, future studies with the WISE database will enable these data to contribute to our understanding of the frequency of DSRs at white dwarfs.  

\section{Optical Detectability of Transits of Dyson Spheres at white dwarf stars}

As noted in the Introduction, a  DSR, or part of it, might be detected as it transits between its star and a space-based telescope such as TESS or Kepler.  In the present section we consider the relative  sensitivity of the optical transit and infrared excess detection techniques.   As noted in Section 4, the current infrared sensitivity limit -- excess IR luminosity due to a DSR divided by bolometric luminosity -- is about 0.1\%.  This limit depends on total solid angle blocked by the entire entourage of artificial constructs as seen from the surface of the white dwarf, but with no constraints on the size of individual structures.  In contrast, the minimal requirement for detection by an optical transit depends on the size of the largest construct, while the total solid angle blocked by the DSR can be far less than 0.1\%. We ask whether a DSR might be detected via transits with Kepler K2 (Howell et al 2014) or with TESS (Ricker et al. 2015).


van Sluijs \& Van Eylen (2018) investigated the sensitivity of K2 to substellar objects that transit white dwarfs.   K2 targets were observed with either long (30 min) or short (1 min) exposure times; about 1000 non-composite, confirmed, white dwarfs were observed with the long cadence mode and about 300 with the short.   For a 7000 K white dwarf (cooling time to reach this temperature $\sim$1.5 Gyr (Dufour et al. 2017), a black body with an orbital semi-major axis equal to a solar radius would have a temperature of 500 K.  The transit time of a small object with this semi-major axis is about 1 min.  So, for a transiting object of given size the detection efficiency is larger in the short cadence mode (see Figure 2 in van Sluijs \& Van Eylen (2018).  The diameters of the smallest objects for which the detection efficiencies are as large as $\sim$10\% are comparable to that of Ceres ($\sim$1000 km).  If a structure has an orbital semi-major axis of a solar radius, then the probability that its orbit would give rise to a transit as
seen from Earth would be $\sim$0.01.  If there are "n" large structures that lie in sufficiently different planes, then the transit probability would increase by a factor of n. 

If a large construct were closer to the white dwarf than a solar radius, either hotter or with high albedo (light shields), then the transit probability could be a few percent.  If the white dwarf were cooler than 7000 K then the orbital semi-major axes of constructs of a given temperature could be even smaller.  But for a cooling time of $\sim$6 Gyr the white dwarf would still be $\sim$5000 K, and thus a transit probability only a factor of two larger for a structure of a given temperature.

To have been detected with K2, a DSR would have to include a large structure (diameter $>$1000 km) with an appropriate orbital inclination with respect to our line of sight.  No such transiting objects were found with K2.  

No summary of TESS studies of white dwarfs is yet available in the literature. TESS observes a given region of the sky for $\sim$27 days with a number of cadences.   Andrew Vanderburg
(personal communication, 2022)  is leading a collaboration that should eventually observe, with 2 minute cadence, about 10,000 white dwarfs; to date about 5000 have been observed.   
Because of TESS' large (21'') pixels, background objects in addition to the white dwarf of interest are often included in the white dwarf pixel.  After correction for the presence of such objects in the white dwarf pixel, the scatter in the measured flux is typically about 10\% for a white dwarf with Gaia G magnitude = 16 (A. Vanderburg, personal communication 2022).  Based on statistics given in Gentile-Fusillo et al (2021) and in Dufour et al (2017) one anticipates about 2500 white dwarfs will be this bright or brighter.  With a representative 3 hour orbital period, a given structure will transit a few 100 times in 27 days, after which one can construct a final phase folded light curve with flux scatter of about 1\% for each of the 2500 white dwarfs.  Thus, a signal-to-noise ratio of, say, 10 in a 27 day average would require a transit depth of $\sim$10\%.  As with K2, such a deep transit requires DSR constucts with diameter of order 1000 km. 
The full TESS sample of 10,000 white dwarfs should contain $\sim$800 with G mag = 15 or brighter, so for these brightest stars a transit depth of a few \% for detection.   Ultimately, when the final TESS white dwarf sample is observed and analyzed, the detection limits quoted in this paragraph may turn out to be too conservative -- because a significant fraction of the final sample should eventually be observed with 20 sec cadence, and/or for longer than 27 days (specifically multiples of 27 by a factor of 2 or more) (A. Vanderburg, personal communication 2022). 

As in Section 4, it is possible to derive constraints on the value of N with use of the Kepler K2 and TESS databases (when the latter becomes available).  However, such constraints will apply only to DSRs that include at least one structure with diameter $\sim$1000 km and will be sensitive to the number of such constructs and the inclination of their orbital planes relative to our line of sight.  Should transits with small fractional dips be detected, it might be possible to distinguish natural from artificial objects by study of the transit shape (e.g., Arnold 2005; Sandford \& Kipping 2019).  However, this would likely require an additional factor of $\sim$10 in signal-to-noise ratio (G. Marcy, personal communication 2022).  
 
\section{The frequency of technological civilizations at main sequence stars}

Evidence of ETI at main sequence stars can be in the form of excess IR emission from a
DSR or as radiation beamed toward Earth at radio, IR, or optical wavelengths for the
purpose of communication.  Each technique has its own advantages and disadvantages.
Concerning beamed transmissions, at radio wavelengths one 
representative major project is ``SETI Observations of Exoplanets with the Allen Telescope Array'' (Harp et al 2016).   A representative search at optical wavelengths would be ``A Search for Laser Emission with Megawatt Thresholds from 5600 FGKM Stars''  (Tellis \& Marcy 2017).   The website technosearch.seti.org gives a comprehensive list of (perhaps all) radio and optical search programs. 
Based on the discussion in Sections 4 and 7, perhaps it would be worthwhile to devote more time to white dwarfs as targets in such searches.  Searches for beamed transmissions usually assume a signal that is so narrow in frequency space and/or has a timing profile such that it cannot be produced by a ``natural'' source.  One huge advantage for a search for a beamed transmission, compared to an IR or transit search for a DSR, is that once one has ruled out terrestrial interference, there are no false positives to contend with.

Mid-infrared satellites have observed numerous main sequence stars in both pointed
(Spitzer) and scanning mode (WISE).  In all cases two important questions are:
(1) what is the minimum fractional IR luminosity (L$_{IR}$/L*) due to a DSR that can be
detected, and (2) if an IR excess is detected, can it be shown with certainty that it
must be coming from a DSR rather than from some "natural" source such as dust or a
brown dwarf. Main sequence stars are much brighter than white dwars, so many more can be searched
for a DSR in a brightness limited sample (see, e.g., Cotten \& Song 2016). Nonetheless, as outlined in Section 7, in some ways white dwarfs are  better targets.

Wright et al (2014) outline a plan to search with Spitzer and WISE at mid-infrared wavelengths for the waste energy of advanced technological civilizations.   They review the idea of Kardashev Type I, II, and III technological civilizations; these correspond, respectively, to technologies that command planetary, circumstellar and galactic energy sources.  Homo sapiens are a Type I civilization and  DSRs correspond to Type II.  However, the focus of the Wright et al paper is on Type III civilizations that, they say, are ``very rare in the local universe''.  

There are two published studies relevant to the frequency of Type II civilizations (= DSRs): Kennedy \& Wyatt (2013) and Moor et al. (2021).  Kennedy \& Wyatt surveyed $\sim$24,000 A8 to K5 type main sequence stars for excess IR emission in the 12 $\mu$m (W3) channel of WISE.  Their goal was to discover stars that are orbited by warm (T $>$ 200 K) dusty disks with ages of at least 8 Myr.  There is no mention in their paper of DSRs or Kardashev civilizations.  Perhaps this is why the Kennedy \&  Wyatt paper is not cited by Wright et al (2014). 

Only about two dozen warm IR excess stars emerged from this extensive study; many of these were previously known (see Table 1 in Kennedy \& Wyatt (2013).  Important for the present study, none of the Table 1 stars satisfies both the age ($>$4.5 Gyr) and spectral type (F6-G7) of interest to us here, but with the possible exception of HD 154593.  However, from a Gaia proper motion anomaly, Kervella et al (2019) deduce that HD 154593 has a cool unseen companion. C. Melis (personal communication, 2021) has carried out an extensive search of the literature over a much broader range of stellar phase space than covered by Kennedy \& Wyatt for stars that would satisfy our constraints for age and spectral type; Melis found no such stars with a reliable IR excess.

Thus, none of the Kennedy \& Wyatt sample of 24,000 stars 
shows evidence for a DSR; only a fraction of the 24,000 satisfy our age and spectral type constraints.  Their Figure 1 of spectral types suggests that about 12,000 stars are in the range F6-G7.  Their Figure 2 gives the age distribution; $\sim$20\% are older than 4.6 Gyr.    Thus we can say that none of  $\sim$2400 stars with the appropriate age and spectral type show evidence of a DSR.
Similar to the white dwarfs discussed in Section 4, the absence of a DSR at these 2400 main sequence stars is for fractional IR luminosities --  excess IR luminosity divided by bolometric luminosity -- of about 0.1\% or greater.  In both classes of target stars one is dealing with a brightness limited sample.  But even though the white dwarfs are fainter than the main sequence stars, for the former sample the IR data are from Spitzer, in contrast to the less sensitive WISE data used for the main sequence sample.

From Equation 3 we see that about 3 billion F6 through G7 stars are now on the main sequence.  As noted in Section 4, studies of Kepler and other data bases by Zink \& Hansen (2019) and by Bryson et al. (2021) suggest that about 30\% of G-type stars have a potentially habitable planet, or about 10$^9$ such planets.  If a million (10$^6$) of these evolve to a technological civilization that constructs a DSR, then one should have been found in the Kennedy \& Wyatt sample of 2400 stars.   Thus, we can say that at most about one in a 1000 G-type stars with habitable planets harbors a planet with life that eventually evolves to a state where it constructs a DSR with luminosity at least 0.1\% that of its host star.

It would be worthwhile to analyze the Moor et al. (2021) database for DSRs.  But this is not possible to do without more detailed information on stellar spectral types and age than is presented in their paper.  We can, however, say that none of the warm IR excess stars identified by Moor et al appears to be a good candidate for a DSR.

The above discussion and that in Section 4 assume that all N civilizations arise independently and do not colonize other stars.   Hansen \& Zuckerman (2021) argue that many such migrations will be motivated by stellar evolution and will be to M-type stars.  If so, then M stars would represent a plausible host for DSRs.  However, the false positive rate for excess IR emission at M stars that are cross correlated with the WISE database is large (Silverberg et al 2018) and, at any rate, no one has ever confidentally identified an M star of age $\sim$5 Gyr or greater that possesses an IR excess.



\section{Discussion}

In Sections 4 and 6 we appraised the occurrence frequency of DSRs at white dwarfs and at main sequence stars based on various observations and assumptions.  Here we compare the relative advantages and shortcomings involved when white dwarfs or main sequence stars are utilized as determinants of DSR frequency.   
The IR surveys referred to in Sections 4 and 6, have similar sensitivities to DSRs ($\sim$0.1\%) when the luminosity of a DSR is measured as a fraction of the bolometric luminosity of a central star; we will call this fraction $\tau$.  At this point in time, no evidence exists of a DSR of this fractional brightness or larger at any of the surveyed white dwarfs and main sequence stars.

No doubt, motivation is a major factor that determines whether an advanced technological civilization would want to construct a DSR. As noted in the Introduction, coping with stellar evolution is one unavoidable motivation.  Interior planets will be destroyed during the giant (AGB) phase of stellar evolution, while surviving outer planets will become exceedingly cold as the white dwarf cools.  Thus, arguably, eventually all organisms will have migrated from planetary surfaces to artificial space colonies. If so, then any civilization that orbits a white dwarf must produce a DSR.  A question then is, what is a likely value for $\tau$?

One can envision reasons why $\tau$ at a white dwarf would be quite large.
Far more organisms can live in an ensemble of space colonies than can live on a planet such as Earth.  Thus, the value of $\tau$ depends on the numbers of organisms, how much space each one would desire, and how much energy would be required per capita.  Needless to say, from our youthful, naive, vantage point we simply don't know the answers to such questions.  About all that one might say is that most Homo sapiens would surely prefer to live in a huge space colony, the bigger the better.  If this motivation is retained even in an advanced long-lived civilization, then $\tau$ could be quite large; easily 0.1\%.  And, as noted in Section 5, really large constructs might produce detectable transit signals.  


Thus, $\tau$ = 0.1\% would seem to be a significant search sensitivity, and upper limits to the frequency of DSRs with this luminosity at white dwarfs, meaningful.   Unfortunately, given their low apparent brightness, as noted in Section 4, only of order 100 white dwarfs with main sequence progenitors in the appropriate mass range were examined with Spitzer.  Given this small sample size, we estimated that at least 3\% of all habitable planets at G-type main sequence stars must spawn life that eventually evolves to technology in order for at least one to have been detected with Spitzer at a white dwarf.  This seems a bit ``optimistic''.

Because a few 1000 G-type main sequence stars with appropriate age and spectral type were examined with WISE, as noted in Section 6, only one in a 1000 or so G-type stars needs to spawn life that eventually develops technology for a DSR with $\tau$$>$0.1\% to have been detected in the Kennedy \& Wyatt (2013) survey.  However, main sequence stars suffer at least two disadvantages as target stars when compared to white dwarfs;  one disadvantage would be less motivation to build a substantial DSR because one's home planet remains a good abode for life.  In our own solar system, if a sunshield is constructed and employed at the inner Earth-Sun Lagrange point -- to counter the increasing solar luminosity -- then Earth could remain quite habitable for a few Gyr more into the future.  Perhaps a more important consideration would be the greater luminosity, say about a factor of 1000, of the Sun compared to a typical white dwarf.  For a DSR with the same $\tau$ and temperature around the Sun as one around a white dwarf, a DSR at the former would have to have 1000 times the area of one at the latter.  While there is sufficient material in the asteroid belt to build such an extensive DSR, would the motivation to do so exist?  For transits of main sequnce stars by structures with temperatures in the range 300 to 1000 K, the orbital period would be much longer than around white dwarfs, thus relatively few transits per year. For a given structure, the probability of proper alignment of orbital plane and line of sight to Earth would be small and its required cross section would be larger than that of Ceres.

\section{CONCLUSIONS}  

We have considered the detection of a Dyson Sphere or Ring (DSR) at a white dwarf star via its infrared emission or via a transit between our telescopes and the star.  Of order 100 white dwarfs of appropriate mass were observed in the infrared with the Spitzer Space Telescope; no plausible DSR candidate has been identified.  We also considered DSRs at main sequence stars; a few 1000 of appropriate age were observed with the Wide-field Infrared Survey Explorer (WISE) and no plausible DSR candidate was identified.  These results, along with a few reasonable assumptions, can be used to place limits on the number of technological civilizations in the Galaxy or, alternatively, on the fraction of habitable planets that orbit solar type stars  and on which a technological civilization eventually emerges and subsequently constructs a DSR.  

We discussed transit surveys of white dwarfs with the Kepler 2 and TESS missions. These could, in principle, detect a DSR that includes at least one structure with diameter $\sim$1000 km.  Detection of a DSR via its infrared excess emission depends on the total solid angle blocked by the entire entourage of artificial constructs as seen from the surface of the white dwarf, but does not require the existence of any really large individual structure.  In contrast, the minimal requirement for detection by an optical transit depends on the size of the largest construct, while the total solid angle blocked by a DSR can be far less than is required for detection of an infrared excess.
  
Both the white dwarf and main sequence studies were sensitive to DSRs whose infrared luminosity ``$\tau$'' is about 0.1\% that of the bolometric luminosity of their central star.
Regarding the number of number ``N'' of technological civilizations in the Milky Way Galaxy, based on the white dwarf studies, if all N originate at solar-type main sequence stars and construct DSRs with $\tau$ at least as large as 0.1\%, then N $<$100 million.  If technological civilizations emerge with equal probability around all stars of solar mass or less, then N$<$$10^9$.  

An alternative way to interpret the observational results is to ask: what fraction of potentially habitable planets that orbit solar-type stars spawn living organisms that eventually evolve to a technology that then constructs a DSR with $\tau$ at least as large as 0.1\%?  From the white dwarf studies we estimate that this fraction is at most 3\%.  From the main sequence studies the upper limit to this fraction could be as small as 0.1\%, but with a few potentially significant caveats.  Additional examination of the existing databases would be worthwhile to ensure that no DSR was missed.  

Substantial progress to improve the current limits, or to detect a DSR, will require a new mid-infrared space telescope.  The diameters of WISE, Spitzer, and JWST are 40, 85, and 650 cm, respectively.   JWST with its long slew time and numerous Galactic and extragalactic objects waiting in line to be observed is not the right telescope with which to search for DSRs.  

The median Gaia G-magnitude of the Roccetto et al (2015) white dwarf sample was 15.4.  According to Gentillo-Fusillo et al (2019), there are about 5000 white dwarfs within 200 pc of Earth and brighter than magnitude 17.  Given that about 25\% of white dwarfs have masses in the (low mass) range of relevance for the possible existence of a DSR (as discussed in Section 4), a scanning telescope operated in a similar way to WISE, but with the diameter of Spitzer, could improve by an order of magnitude the limits on DSR frequency derived in this paper.   Such a telescope would yield much new science in general and would not be too costly.   The 30-m class ground-based telescopes currently envisioned or in construction could be used to vet any potential DSR candidates.     
  
Future telescopes, such as LSST, may be suitable for detection of transits of DSRs at white dwarfs.  But individual constructs would have to have diameters $\sim$1000 km, i.e. comparable to that of Ceres
(Cortés \& Kipping 2019).  ESA's planned PLATO spacecraft, if successful, should be at least a few times more senstitive than TESS for detection of very shallow transits at white dwarfs and thus able to detect constructs with diameters a few times smaller than 1000 km.
  

\acknowledgements

The author is grateful to Andrew Vanderburg and Beth Klein for a variety of assistance, to referee Geoffrey Marcy for many useful suggestions, and to Carl Melis, Jay Farihi, Jill Tarter, and Brad Hansen for helpful advice.  This research was supported in part by grants to UCLA from NASA and the NSF.  We have made use of NASA's Astrophysics Data System.


\vskip 0.2in

DATA AVAILABILITY

No new data were generated or analysed in support of this research.



\end{document}